# Event-controlled constructions of random fields of maxima with non-max-stable dependence

Mathias Raschke

*Stolze-Schrey-Str.1, 65195 Wiesbaden, Germany*

Phone +49  611 98819561

E-Mail MathiasRaschke@t-online.de

**Abstract:** Max-stable random fields can be constructed according to Schlather (2002) with a random function or a stationary process and a kind of random event magnitude. These are applied for the modelling of natural hazards. We simply extend these event-controlled constructions to random fields of maxima with non-max-stable dependence structure (copula). The theory for the variant with a stationary process is obvious; the parameter(s) of its correlation function is/are determined by the event magnitude. The introduced variant with random functions can only be researched numerically. The scaling of the random function is exponentially determined by the event magnitude. The location parameter of the Gumbel margins depends only on this exponential function in the researched examples; the scale parameter of the margins is normalized. In addition, we propose a method for the parameter estimation for such constructions by using Kendall's τ. The spatial dependence in relation to the block size is considered therein. Finally, we briefly discuss some issues like the sampling.

## 1. Introduction

Max-stable random fields can be constructed according to the first two theorems of Schlather (2002). These are very attractive from the practical point of view because they include a random variable that is a kind of event magnitude and the spatial pattern of the local event impact is modelled explicitly by a random function or a stationary random process. In contrast, Wadsworth and Tawn (2012) formulate an inverted max-stable field with direct transformation of a max-stable random field. There are important earlier works (e.g. de Hahn, 1984; Smith, 1990) and recent papers (e.g., Kabluchko et al., 2010; Engelke et al., 2011; Hoffmann, 2013; Robert, 2013) about max-stable random fields. Besides, max-stable random fields are already applied for the modelling of natural hazards such as rainfall (Coles, 1993; Davison, 2012). Earthquake hazard models are similar to the max-stable random fields and include a physical definition of the magnitude (Raschke, 2013). However, not all natural hazards have a max-stable dependence structure (copula). For example, river floods have no max-stability in the spatial dependencies between discharge peaks of different gauging stations because there is no asymptotic tail dependence between the discharges (Keef et al., 2009a, 2009b). This is the motivation to extend the event-controlled construction of max-stable random fields to random fields of maxima with non-max-stable dependence and with the possibility to control the behaviour of the dependence in relation to the block size.

We explain our extension for the variant with a stationary process in the next section. Then we construct a maxima field with the random function that depends on event magnitude and research it numerically with examples. In section 4 we suggest an estimation method by using Kendall's $\tau$. At the end, we conclude the results and discuss briefly some issues like the sampling.

## 2. The construction with a stationary random process

The proved Theorem 2 of Schlather (2002) can be also written as

Theorem 1: *Let Y be a stationary process on $\mathbb{R}^d$ with expectation $\mathbb{E}exp(Y(o))=1$, and let $\Pi$ be a Poisson process on (-∞,∞) with intensity measure $d\Lambda(m)=exp(-m)dm$. Then*

$$Z(x) = \max_{m \in \Pi}(m + Y_m(x))$$

*is a stationary max-stable process with unit Gumbel margins, where the $Y_m$ are i.i.d. copies of Y for all m in (-∞,∞).*

Therein our maxima are the logarithmized version of Schlather's maxima. The random event magnitude is $m$. We can simply modify it:

Theorem 2: *Let Y be a stationary process on $\mathbb{R}^d$ with expectation $\mathbb{E}exp(Y(o))=1$, and let $\Pi$ be a Poisson process on (-∞,∞) with intensity measure $d\Lambda(m)=exp(-m)dm$. Then*

$$Z(x) = \max_{m \in \Pi}(m + Y_m(x))$$

*is a stationary process of maxima with unit Gumbel margins, where the $Y_m$ are i.i.d. copies of Y for all m in $(\infty,\infty)$ except the dependence between $Y_m(x_1)$ and $Y_m(x_2)$ that depends on m.*

The proof is trivial because the generating of $Z(x)$ for a fixed point $x$ is the same as in Theorem 1: the maxima of sums of two random variables. □

Max-stability according to the definition of Schlather (2002) is only the special case in which $m$ has no influence on $Y_m$. In the further research, we consider the block maximum

$$Z_k(x) = \left(\max_{m \in \Pi} {}_1(m + Y_m(x)), \max_{m \in \Pi} {}_2(m + Y_m(x)), ..., \max_{m \in \Pi} {}_k(m + Y_m(x))\right) \tag{1}$$

at the point $x$ with block size $k$. The theorem 1 and 2 applie for $k=1$. But the other corresponding block maxima are also Gumbel distributed with cumulative distribution function

$$G(x) = \exp(-\exp(-(x - a)/b)), \tag{2}$$

wherein the scale parameter is normalized with $a=1$ and the location parameter is $b=\log(k)$ .

Now we present some examples for our construction with a stationary Gauss process. The Gauss process is parameterized by the stationary variance $Var(Y)$, the stationary expectation $\mathbb{E}Y=-Var(Y)/2$, and a correlation function $\rho(x_1,x_2)$ with parameter(s) that depend on random magnitude $m$. We show in Figure 1a and b the relation between Kendall's $\tau$ and the distance $|x_1 - x_2|$ for some examples for stationary Gauss process in an one dimensional space. These functions also depend on the block size $k$ and are estimated for a sample of $Z_k(x)$ of size $n=50.000$ being Monte Carlo simulated. Figure 1c and 1d depict realizations of such fields of maxima. The larger or smaller spatial dependence of the maxima results in graphs with larger or smaller volatility.

If the dependency between $Z_k(x_1)$ and $Z_k(x_2)$ decreases with increasing $k$ then the limit state can be a white noise field for $Y(x)$. Its variance determines then the dependencies between $Z_k(x_1)$ and $Z_k(x_2)$. The limit state for increasing dependency with increasing $k$ can be the fully dependence. These are only the limits of possible limit states of this flexible approach.

The hypothesis of unit Gumbel distribution for $Z_1(x)$ for $k=1$ has been tested for at least one sample size $n=50.000$ for each researched variant by the Anderson-Darling test for the fully specified distribution according to Stephens (1986). The number of rejections of distribution hypothesis for a significance level of $\alpha=5\%$ corresponds with the significance level (type 1 error).

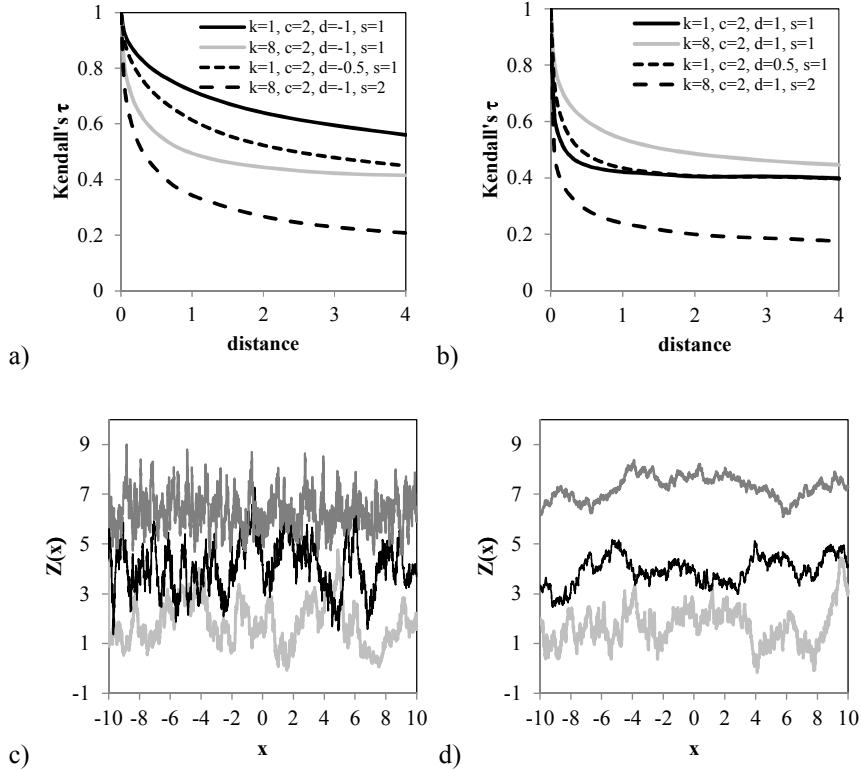

Figure 1: Examples for constructions according to Theorem 2 in $\mathbb{R}^1$ with a stationary Gauss field with correlation function $\rho(x_1, x_2) = \exp(-|x_1 - x_2|/\exp(d(m - c)))$: a) decreasing $\tau$ for increasing $k$, b) increasing $\tau$ for increasing $k$, c) realizations for standard derivation $s=1$ and $d=0.5$, $c=2$, $k=1$ (light gray line), $k=8$ (black line) and $k=64$ (dark gray line), d) like c) but with $d=-0.5$

## 3. The construction with a random function

The proved theorem 1 of Schlather (2002) can be also written as

Theorem 3: *Let $Y$ be a measurable random function with expectation $\mathbb{E}\int_{\mathbb{R}^d} \exp(Y(x))\, dx = 1$. Let $\Pi$ be a Poisson process on $\mathbb{R}^d \times (\infty, \infty)$ with intensity measure $d\Lambda(y, m) = dy\, \exp(-m) dm$, and $Y_{y,m}$ i.i.d. copies of Y; then*

$$Z(x) = \max_{(y,m) \in \Pi} (s + Y_{y,m}(x - y))$$

*is a stationary max-stable process with unit Gumbel margins.*

Therein our maxima are the logarithmized version of Schlather's maxima and the random event magnitude is again $m$. We could not formulate and prove a theorem similar to our Theorem 2. But we could find constructions similar to Theorem 3 by heuristic research. Therein, we use probability distribution functions $f(x)=exp(Y(x))$. These are centred (expectation is 0), and their standard derivation $s$ is determined by $m$ with

$$s = \exp(d(m - c)) \, . \tag{3}$$

Max-stability of the dependence is the special case with $d$=0. We show in Figure 2a the relation between Kendall's $\tau$ and the distance $|x_1\text{-}x_2|$ for some examples with a Gauss distribution, a Laplace distribution and a uniform distribution. The scale parameter of the resulting Gumbel margins according to Eq.(2) is $b$=1. However, the location parameter of the Gumbel margins is not necessarily $a$=0 for $Z_1(x)$. We present some results in Tab.1. An interesting fact is that the differences between the estimated location parameters $\hat{a}$ (Eq.(2)) of all variants of distribution functions for $Y(x)$ are not significant; they correspond to the standard error of the estimation. We can state that the location parameter depends only on the function $m{\to}s$ in our examples. Additionally, we have tested positive that the margins are Gumbel distributed with the Anderson-Darling test of Stephens (1986) for known scale parameter $b$=1 and estimated location parameter $\hat{a}$. We show further examples in Figure 2b with a normal distribution for $Y(x)$ and realizations in Fig.2c and d. Once again, we have computed all functions by using Monte Carle simulations according to the notes of Schlather (2002). An acceptable Monte Carlo simulation is difficult or impossible for the case that the scaling $s$ of $Y(x)$ is increasing for increasing random magnitude $m$ because the considered range of the space needs to be very large or infinite for a sufficient consideration of large magnitudes.

There are further opportunities. Different distribution functions $f(x)$ could be mixed for $Y(x)$. They could be also used with a stationary random field $W(x)$ with expectation $\mathbb{E}exp(W(x))$=1 in $Y(x)=log(f(x))+W(x)$. This would be only a special case of a random function. We could also apply a random parameter $s$ and determine only its expectation by Eq.(3).

Tab.1: Estimation results for the Gumbel distribution according to Eq.(2) for $Z_1(x)$ being generated by construction with a random function with scaling according to Eq.(3) (parameters in the first/second row; critical values for the Anderson-Darling test are 1.321 for $\alpha$=5% and 1.062 for $\alpha$=10% [known scaling parameter]; sample size $n$=50.000)

| Value | Gauss distribution | | Laplace distribution | | Uniform distribution | |
|---|---|---|---|---|---|---|
| | $c$=2, $d$=-0.3 | $c$=3, $d$=-0.2 | $c$=2, $d$=-0.3 | $c$=3, $d$=-0.2 | $c$=2, $d$=-0.3 | $c$=3, $d$=-0.2 |
| $a$ of Eq.(2) with $b$=1, standard error ±0.00811 | -0.1884 | -0.2678 | -0.1900 | -0.2530 | -0.1912 | -0.2655 |
| Anderson-Darling test statistic | 0.8646 | 0.5099 | 0.5622 | 0.9319 | 0.3818 | 0.7810 |

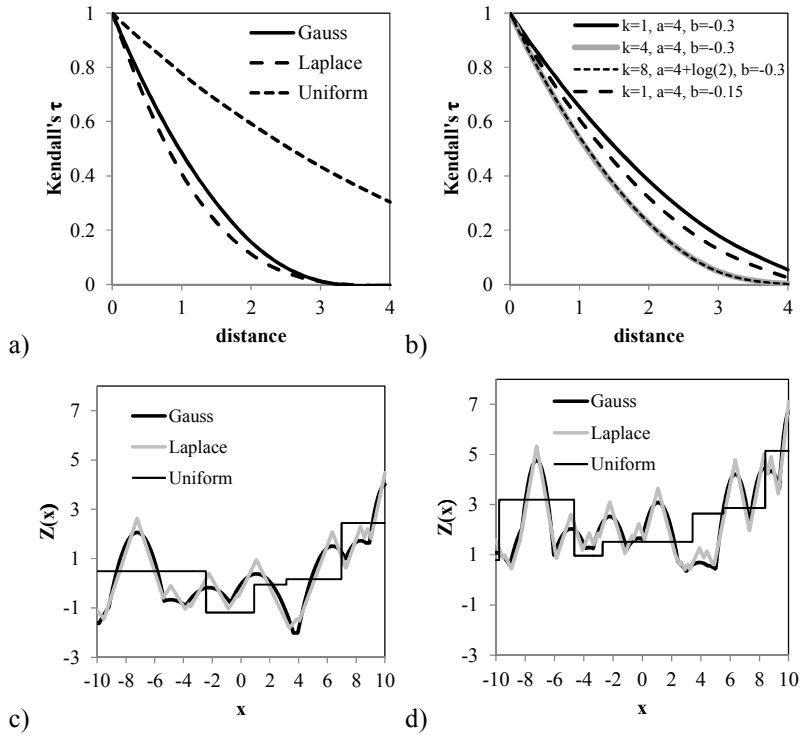

Figure 2: Examples for constructions in $\mathbb{R}^1$ according to Theorem 3 but without max-stable dependence: a) relation between distance $|x_1\text{-}x_2|$ and Kendall's $\tau$ for different distribution functions ($k$=1, Eq.(3) with $d$=-0.3, $c$=2), b) for a Gauss distribution with different variants of parameterization, c) realizations for $k$=1, $d$=-0.3 and $c$=2, d) like c) but for $k$=8 (The same series of random numbers are used for each realization to provide a better comparison.)

# 4. An estimation approach

The maximum composite likelihood estimator is applied already for the inference of max-stable random fields (e.g., Ribatet et al., 2012; Davison et al., 2013). We also use this approach. The parameters $\theta_y$ of the random process $Y(x)$ or the random function $Y(x)$ respectively the parameters $\theta_m$ of their depending on the random event magnitude $m$, e.g. a function $m{\rightarrow}\theta_y$, can be estimated by maximization of the following composite logarithm likelihood function (pairwise likelihood)

$$l(\theta_y, \theta_m) = \sum_{k=1}^{k_{max}} \sum_{i \neq j} log\left( \phi\left( \frac{\hat{\tau}_{obs,i,j,k} - \hat{\tau}_{simu,i,j,k}}{\sqrt{Var(\hat{\tau}_{obs,i,j,k}) + Var(\hat{\tau}_{simu,i,j,k})}} \right) \right) \qquad (4)$$

with the probability density function $\phi$ of the standard Gauss distribution. Kendal's $\hat{\tau}_{obs,i,j,k}$ is the estimation with observations of $Z_k(x_i)$ and $Z_k(x_j)$ according to Kendall (1938) and $\hat{\tau}_{simu,i,j,k}$ is the estimation with values of $Z_k(x_i)$ and $Z_k(x_j)$ being simulated with the parameters $\theta_y$ and $\theta_m$ for block size $k$. The estimation variance $Var(\hat{\tau})$ is also given by Kendall (1938). We can only estimate Kendal's $\tau$ for simulated values because we have no expression for the function $(x_i, x_j){\rightarrow}\tau$. But $Var(\hat{\tau}_{simu,i,j}$ tends to be nil with increasing number of simulations. The Eq.(4) bases on the construction of confidence intervals for Kendall's $\tau$ and was already applied by Raschke et al. (2011) for the estimation with $Z_l(x)$ for a storm model of Switzerland. Here we have to consider Kendall's $\tau$ for $k>1$ to control the behaviour of the dependence in relation to the block size $k$. The block size can be defined by the length of observation period, e.g. the annual maxima or the maxima of two or four years.

We draw attention to the fact that the concrete model has to be built carefully to ensure identifiability. Furthermore, the computational burden can be reduced for the case that $Z(x)$ is isotropic. The function $|.| \rightarrow \tau$ could be computed by Monte Carlo simulations for a relative small number of defined values of distance $|.|$ and the concrete value of $\hat{\tau}_{simu,j}$ can then be interpolated for $|x_i\text{-}x_j|$.

# 5. Conclusion and discussion

In this note, we have suggested two variants of flexible event-controlled random fields of maxima that can be non-max-stable. These are based on the construction of max-stable random fields according to Schalther's Theorems. Max-stability of the dependence is just a special case of our constructions. We have also suggested an inference method. The explicit generation of the events can be important for the practical application of the model because the risk is determined by the losses and damages of a concrete event. Infrastructure systems have to resist a concrete event (e.g., Raschke et al., 2011) and excess of loss reinsurance contracts covers the aggregated loss of single events (e.g., Eling and Toplek, 2009).

It would be very helpful if a general theorem were formulated and proved for the variant with random functions (section 3). Explicit expressions for the functions $(x_i,x_j) \rightarrow \tau$ and the corresponding copulas would also be a large advantage. Nevertheless our approach is already an improvement because it is very flexible. This is important for practical applications because models with known copula expressions could differ considerably from observations (e.g., Davison, 2012, Fig.6 and 9). There is also the question if the concrete physical phenomenon is max-stable. A goodness-of-fit test for a copula could validate such an assumption but it needs a relatively large sample size for its powerful application (Genest et al., 2009), this applies especially for max-stable copulas (Genets et al., 2011). A large sample of annual maxima is more seldom in practice (e.g., Raschke et al., 2011). Daily observations are frequently not a better sample because of the following reasons: the original variables such as wind speed or river discharge of two points in the geographic space are not synchronized. There can be an unknown, non-stationary time lag between the local event maxima (e.g., Appel, 1983). Moreover, the observations at one point are not independent in time and frequently not identical distributed (e.g., Coles, 2001, chapter 6). The latter can also apply to observations over a large threshold (e.g., Raschke, 2012, Fig.4). These also limit the practical application of measures for the tail dependency. The comparison of Kendall's $\tau$ for different block sizes $k$ is more robust in terms of the sampling. Future results in field of statistics could lead to a better modelling of natural hazards and risks.

# References

Appel, D.H. 1983. Traveltimes of flood waves on the New River between Hinton and Hawk West, West Virginia. U.S. Geological Survey Water-Supply Paper 2225, United States Government printing office.

Coles, S.G. 1993. Regional Modelling of Extreme Storms via Max-Stable Processes. Journal of the Royal Statistical Society - B 55: 797-816.

Coles, S.G. 2001. An introduction to statistical modeling of extreme values. Springer Series in Statistics, Springer, London.

Davison, A. C., Padoan, S. A., Ribatet, M. 2012. Statistical Modelling of Spatial Extremes. Statistical Science 27: 161-186.

Davison, A.C:, Huser, R., Thibaud, E. 2013. Geostatistics of dependent and asymptotically independent extremes. Math Geosci 45:511–529.

de Haan, L. 1984. A spectral representation for max-stable processes. Ann Probab 12:1194–1204

Eling, M. Toplek, D. 2009. Risk and return of reinsurance contracts under copula models. The European Journal of Finance 15:751 — 775.

Engelke, S., Kabluchko, Z., Schlather, M. 201. An equivalent representation of the Brown–Resnick process. Statistics & Probability Letters 81:1150–1154.

Genest, C., Rémillardb, B., Beaudoinc, D. 2009. Goodness-of-fit tests for copulas: A review and a power study. Insurance: Mathematics and Economics 44: 199–213.

Genest, C., Kojadinivic, I., Slehove, J. 2011. A goodness-of-fit test for bivariate extreme-value copulas. Bernoulli 17:253–275.

Hoffmann M. 2013. On the hitting probability of max-stable processes. Statistics & Probability Letters 83:2516-2521.

Kabluchko, Z., Schlather, M., de Haan, L. 2009. Stationary Max-Stable Fields Associated to Negative Definite Functions. The Annals of Probability 37:2042-2065.

Keef, C., Svensson, C., Tawn, J.A. 2009a. Spatial dependence in extreme river flows and precipitation for Great Britain. Journal of Hydrology 378:240–252.

Keef, C, Tawn, J.A., Svensson, C. 2009b. Spatial risk assessment for extreme river flows. Appl. Statist. 58:601-618.

Kendall, M.G. 1938. A new measure of rank correlation. Biometrika Trust 30:81-93.

Raschke, M., Bilis, V., Kröger, W. 2011. Vulnerability of the Swiss electric power grid against natural hazards. In Proceedings: 11th International Conference on Applications of Statistics and Probability in Civil Engineering (ICASP11), Zurich, Switzerland.

Raschke, M. 2012. Möglichkeiten der mathematischen Statistik zur Schätzung der Hochwasserwahrscheinlichkeit (German, Possibilities of mathematical statistics to estimate flood probability). Wasser und Abfall 06/2012.

Raschke, M. 2013. Statistical modelling of ground motion relations for seismic hazard analysis. Journal of Seismology 17:1157-1182.

Ribatet, M., Cooley, D., Davison, A.C. 2012. Bayesian inference from composite likelihoods, with an application to spatial extremes. Statistica Sinica 22:813-845.

Robert, C. 2013. Some new classes of stationary max-stable random fields. Statistics & Probability Letters 83:1496–1503.

Schlather, M 2002. Models of stationary max-stable random fields. Extreme 5:33-44.

Smith, R. L. 1990. Max-stable processes and spatial extremes. Unpublished

Stephens, M.A. 1986. Test based on EDF statistics. In: D'Augustino RB, Stephens MA (eds) Goodness-of-fit techniques. Statistics: textbooks and monographs, vol. 68. Marcel Dekker, NewYork.

Wadsworth, J.L., Tawn, J.A. 2012. Dependence modelling for spatial extremes. Biometrika 99:253–272.